\documentclass[sigconf]{acmart}
\AtBeginDocument{%
  }

\copyrightyear{2025}
\acmYear{2025}
\setcopyright{acmlicensed}
\acmConference[KDD '25]{Proceedings of the 31st ACM SIGKDD Conference on Knowledge Discovery and Data Mining V.1}{August 3--7, 2025}{Toronto, ON, Canada}
\acmBooktitle{Proceedings of the 31st ACM SIGKDD Conference on Knowledge Discovery and Data Mining V.1 (KDD '25), August 3--7, 2025, Toronto, ON, Canada}
\acmDOI{10.1145/3690624.3709171}
\acmISBN{979-8-4007-1245-6/25/08}

\usepackage{soul}
\usepackage{amsmath}
\usepackage{amsfonts}
\usepackage{amsthm}
\usepackage{booktabs}
\usepackage{xspace}
\usepackage{subfigure}
\usepackage{pifont}
\usepackage{mathtools}
\usepackage{environ}
\usepackage{multirow}
\usepackage{pdfpages}
\usepackage{balance} 
\usepackage{url}
\usepackage{multirow}
\usepackage{caption}
\usepackage{graphicx}
\usepackage{float}
\usepackage{subcaption}

\usepackage{algorithm}
\usepackage{algorithmic}

\newcommand{\eg}{\emph{e.g.,}\xspace}

\begin{document}

\title{MobileSteward: Integrating Multiple App-Oriented Agents with Self-Evolution to Automate Cross-App Instructions}

\author{Yuxuan Liu}\authornote{This work was done when Yuxuan interned at Xiaomi AI Lab.}
\author{Hongda Sun}
\affiliation{%
  \institution{ Gaoling School of Artificial Intelligence, \\ Renmin University of China}
  \city{Beijing}
  \country{China}
}
\email{yuxuanliu@ruc.edu.cn}
\email{sunhongda98@ruc.edu.cn}


\author{Wei Liu}
\author{Jian Luan}
\affiliation{%
  \institution{Xiaomi AI Lab}
  \city{Beijing}
  \country{China}}
\email{liuwei40@xiaomi.com}
\email{luanjian@xiaomi.com}


\author{Bo Du}
\affiliation{%
  \institution{School of Computer Science, Wuhan University}
  \city{Wuhan}
  \country{China}}
\email{dubo@whu.edu.cn}

\author{Rui Yan}\authornote{Corresponding author: Rui Yan(ruiyan@ruc.edu.cn)}
\affiliation{%
  \institution{ Gaoling School of Artificial Intelligence, \\ Renmin University of China}
  \city{Beijing}
  \country{China}
}
\affiliation{%
  \institution{School of Computer Science, Wuhan University}
  \city{Wuhan}
  \country{China}}
\email{ruiyan@ruc.edu.cn}

\renewcommand{\shortauthors}{Yuxuan Liu, et al.}

\begin{abstract}
Mobile phone agents can assist people in automating daily tasks on their phones, which have emerged as a pivotal research spotlight. 
However, existing procedure-oriented agents struggle with cross-app instructions, due to the following challenges: (1) complex task relationships, (2) diverse app environment, and (3) error propagation and information loss in multi-step execution. 
Drawing inspiration from object-oriented programming principles, we recognize that object-oriented solutions is more suitable for cross-app instruction.
To address these challenges, we propose a self-evolving multi-agent framework named \textbf{MobileSteward}, which integrates multiple app-oriented StaffAgents coordinated by a centralized StewardAgent.
We design three specialized modules in MobileSteward: 
(1) \textit{Dynamic Recruitment} generates a scheduling graph guided by information flow to explicitly associate tasks among apps. 
(2) \textit{Assigned Execution} assigns the task to app-oriented StaffAgents, each equipped with app-specialized expertise to address the diversity between apps.
(3) \textit{Adjusted Evaluation} conducts evaluation to provide reflection tips or deliver key information, which alleviates error propagation and information loss during multi-step execution.
To continuously improve the performance of MobileSteward, we develop a \textit{Memory-based Self-evolution} mechanism, which summarizes the experience from successful execution, to improve the performance of MobileSteward.  
We establish the first English Cross-APP Benchmark (CAPBench) in the real-world environment to evaluate the agents' capabilities of solving complex cross-app instructions. Experimental results demonstrate that MobileSteward achieves the best performance compared to both single-agent and multi-agent frameworks, highlighting the superiority of MobileSteward in better handling user instructions with diverse complexity.
\end{abstract}




\maketitle

\begin{figure}[]
\setlength{\abovecaptionskip}{0.3cm}
\setlength{\belowcaptionskip}{-0.5cm}
\centering
\includegraphics[width=\linewidth]{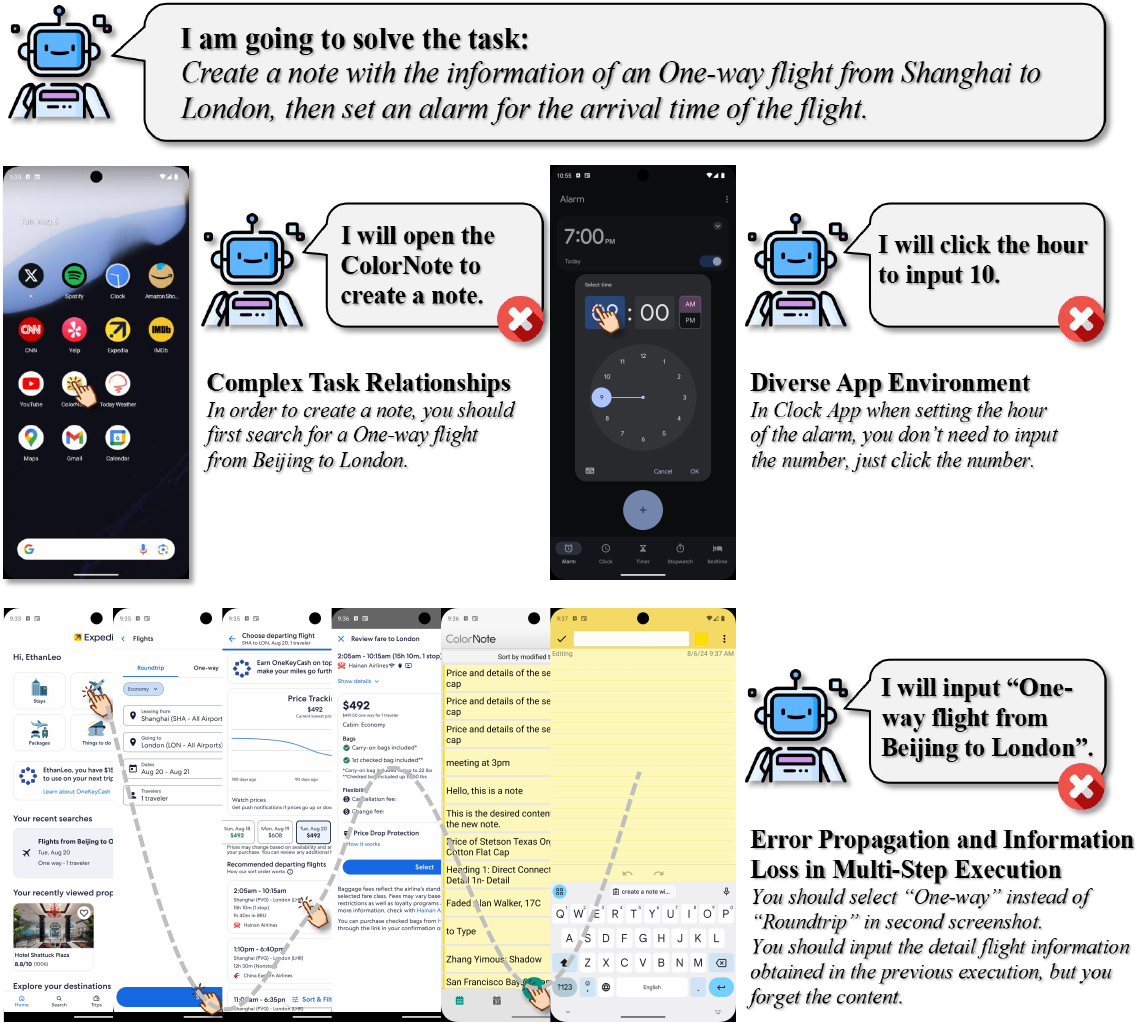}
\caption{Current mobile phone agents suffer from the following challenges when solving cross-app instructions: (1) Complex Task Relationships; (2) Diverse App Environment; (3) Error Propagation and Information Loss in Multi-Step Execution.}
\label{fig:motivation}
\end{figure}

\section{Introduction}

Mobile phone agents can assist people to automate simple tasks and bring much convenience to people's daily lives, which have emerged as a pivotal research spotlight~\cite{wen2023empowering,yang2023appagent,wang2024mobile}. 
The rapid advancement of mobile technology has led to the proliferation of apps with diverse functionalities, allowing users to accomplish increasingly complex tasks. Consequently, developing more powerful agents capable of handling complex user instructions in such environments is of great importance and application prospect.

Existing mobile phone agents have achieved some encouraging results on automated task execution.
Several API-based solutions have been successfully deployed in real-world mobile phones, such as Siri, Google Assistant, and XiaoAI~\cite{wen2023empowering}.
Despite these advances, API access constraints have prompted the exploration of alternative methodologies, including 
simulating user operations by learning the mapping between instructions and action sequences~\cite{li2020mapping, li2021learning}, understanding app page transitions in relation to actions~\cite{lee2023explore,wen2024autodroid,wu2024mobilevlm}, and augmenting agents' decision-making with the structured interface representations and historical states~\cite{zhan2023you,ma2024comprehensive,dorka2024training}. 
With the appealing performance of Large Language Models (LLMs) and Large Multi-modal Models (LMMs), many efforts have leveraged their logical reasoning, role-playing, and image understanding abilities to create more generalized agents~\cite{sun2024harnessing,sun2024facilitating,liu2025bidev,wang2023enabling,yan2023gpt,zheng2024gpt}, which can autonomously analyze user intent, comprehend screen content, and infer the appropriate actions to execute instructions~\cite{hong2024cogagent,baechler2024screenai,yang2023appagent,wang2024mobile}.

However, implementing automated task execution on mobile phones in real-world scenarios remains challenging. As illustrated in Figure~\ref{fig:motivation}, although existing mobile phone agents can solve straightforward tasks within a single app, they struggle with cross-app instructions due to the following challenges: 
(1) \textit{Complex task relationships}: A complex user instruction often contains intricate dependencies between different subtasks. Managing these dependencies and scheduling tasks accurately is challenging for current agents ; 
(2) \textit{Diverse app environment}: Unlike general agents, the mobile environment is highly diverse, with apps varying widely in functionality, content, and user interface design. This diversity complicates the agent's ability to uniformly understand and interact with different apps; 
(3) \textit{Error propagation and information loss in multi-step execution}: Multi-step task execution may cause the propagation of previous errors or the loss of key information, thus disrupting the successful completion of subsequent actions. 
To effectively handle cross-app instructions, it is essential to not only coordinate tasks between apps but also execute actions accurately within apps. Current methods typically rely on a single agent, which lacks the versatility and specialization needed to perform both aspects effectively.

Previous research has demonstrated that the multi-agent framework can effectively address complex tasks, such as program development~\cite{qian2023communicative,hongmetagpt}, game strategies~\cite{zhu2024player,noh2024llms,huang2024far}, or intricate reasoning~\cite{chan2023chateval,sun2024determlr,kim2024can}. 
These methods typically divide the entire task into a multi-stage pipeline to reduce the difficulty of each stage, and then design specialized agents for each stage to complete the task. 
However, research on multi-agent frameworks in mobile phone is still underdeveloped. 
Although MobileAgent-v2~\cite{wang2024mobilev2} follows the procedure-oriented idea to combine the Planning Agent, Decision Agent, and Reflection Agent at each execution step, its performance in solving cross-app instructions is still unsatisfactory.

Inspired by the principles from programming languages like C++, we find that cross-app instruction is more suitable to be solved using object-oriented design solutions~\cite{lieberherr1988object}. 
Given that existing agents can effectively solve simple tasks within a single app, if we can treat the app-specialized agent as the object, we only need to concentrate on how to schedule and assign tasks among these agents. 
Based on the above insights, we propose \textbf{MobileSteward}, a self-evolving multi-agent framework for complex cross-app instructions, which integrates multiple app-oriented StaffAgents coordinated by a centralized StewardAgent to collaboratively solve cross-app instructions.
MobileSteward features three specialized modules: (1) \textit{Dynamic Recruitment}: StewardAgent splits the instruction into app-oriented tasks and recruits related StaffAgents, then establishes the information flow among tasks to generate the scheduling graph. (2) \textit{Assigned Execution}: StaffAgent operates the corresponding apps to complete the assigned tasks using app specialized information and return the execution history. (3) \textit{Adjusted Evaluation}: StewardAgent evaluates the StaffAgent's execution, provides reflection tips for errors or summarizes the results of correct execution to deliver the information and adjust the subsequent schedule.

Similar to the steward in reality, with the accumulation of experience, he will become more familiar with the staff about their suitable work. Therefore, we propose \textit{Memory-based Self-evolution} mechanism to achieve the continuous optimization of the MobileSteward framework through the self-summarization of experience from the successful execution. Specifically, we equip the StewardAgent with a \textit{Staff Expertise Memory} and the StaffAgent with a \textit{Task Guideline Memory}. After StaffAgent successfully completes an assigned task, StewardAgent will summarize the execution process, extract the staff expertise and task guidelines, and update the memory. The successful experience injected by memory assists StewardAgent in accomplishing more accurate Dynamic Recruitment and provides StaffAgent with more effective guidance for Assigned Execution.

To evaluate the effectiveness of our MobileSteward, we propose Cross-APP Benchmark (CAPBench), a more challenging benchmark that is specifically designed for complex cross-app instructions. Each instruction inherently requires the interaction of multiple apps, where the tasks are interrelated across these different apps. 
We compare MobileSteward with existing mobile phone agent baselines and the experimental results demonstrate that our proposed MobileSteward can achieve the best performance on solving cross-app instructions that are challenging for both single-agent and multi-agent baselines. Detailed experimental analysis validates the effectiveness of our proposed modules as well as the self-evolution.

Overall, our main contributions can be summarized as follows:

$\bullet$ We propose \textbf{MobileSteward}, a novel multi-agent collaboration framework based on mobile phone environments, which comprises a centralized \textbf{StewardAgent} and several app-oriented \textbf{StaffAgents}.

$\bullet$ We design three specialized modules: \textit{Dynamic Recruitment} for instructive task scheduling, \textit{Assigned Execution} for accurate task execution and \textit{Adjusted Evaluation} for phased task adjustment.

$\bullet$ We introduce \textbf{Memory-based Self-evolution} to continuously optimize MobileSteward with \textit{Staff Expertise Memory} and \textit{Task Guideline Memory} accumulated from successful execution.

$\bullet$ We construct the first English Cross-APP Benchmark (CAPBench) in the real-world environment. The experimental results demonstrate that MobileSteward achieves the best performance in handling complex cross-app instructions.

\section{Related Work}

\subsection{Automated Task Execution on Mobile Phone}

Mobile phones have become so inseparable from our lives, that the development of automated user task execution on mobile phones has become a research spotlight, which can be categorized into three types of methods:
(1) \textbf{API-based methods}, which are favored by industry and have already been deployed for user in actual mobile phones, \eg Siri, Google Assistant and XiaoAI ~\cite{wen2023empowering}. However, such methods are limited by API access and invocation to some extent.
(2) \textbf{GUI-based methods}, which seek for automated task execution by simulating interactions with the graphical user interfaces (GUIs) ~\cite{li2020mapping, li2021learning, burns2022interactive, sun2022meta}. These methods usually require screen summarizing~\cite{wang2021screen2words, li2021screen2vec, zhang2021screen}, widgets recognition~\cite{li2020widget, chen2022towards} and command grounding~\cite{bai2021uibert, burns2022dataset} to augment the GUI understanding and action prediction. Moreover, Spotlight~\cite{li2023spotlight} designed a Region-of-Interest (ROI) Align to localize to the regions and widgets that more relevant to the task.
(3) \textbf{Experience-based methods}, which learn from the historical experience. MobileGPT~\cite{lee2023explore} constructs a Hierarchical App Memory through exploration and then uses Flexible Task Recall in the execution phase. AutoDroid~\cite{wen2024autodroid} constructs the UI Transition Graph (UTG) by exploring during the offline phase, which in turn is extracted to form memory.

The diversity among apps makes existing methods ineffective in solving tasks on various apps, so we design Assigned Execution to utilize app-oriented StaffAgents to complete tasks on specific apps.

\subsection{LLM/LMM Agents on Mobile Phones}

The rapid development of LLMs/LMMs has encouraged them to be adopted as agents on mobile phones ~\cite{wen2023droidbot}. These methods utilize LLM's powerful semantic reasoning capabilities to analyze the tasks~\cite{venkatesh2022ugif,wang2023enabling} or LMM's excellent image comprehension capabilities to assist the GUI understanding~\cite{yan2023gpt, zheng2024gpt}. We can divide them into three categories:
(1) \textbf{Pre-trained methods}: CogAgent~\cite{hong2024cogagent} and ScreenAI~\cite{baechler2024screenai}, pre-train a visual language model on a mix of screen tasks (QA, summarization, annotation and navigation) to build a general agent for automated task execution.
(2) \textbf{Fine-tuned methods}: These methods usually include the historical information to assist in action decisions. 
Auto-UI~\cite{zhan2023you} introduces historical actions during fine-tuning on a large scale dataset AITW~\cite{rawles2024androidinthewild} and can be improved by adopting Chain-of-Action-Thought(CoAT)~\cite{zhang2024android}. 
CoCo-Agent\cite{ma2024comprehensive} augments the screenshot with the textual layout representation and conduct conditional action prediction. 
(3) \textbf{Inference methods}: These methods instruct LLMs/LMMs for planning, decision making or reflection to automate tasks~\cite{deng2024mobile}.
AppAgent~\cite{yang2023appagent} generates documents by self-exploration/demo-watching and adopts SoM~\cite{yang2023set} to assist in action decision. 
MobileAgent~\cite{wang2024mobile} augments action grounding with visual perception module and action execution with self-planning and self-reflection.

These single-agent methods struggle to solve cross-app instructions because of the long execution, thus we designed Adjusted Evaluation to alleviate the information loss and error propagation.

\subsection{Multi-Agent Framework}

The success of AutoGPT~\cite{autogpt}, HuggingGPT~\cite{shen2024hugginggpt} and OpenAGI~\cite{ge2024openagi} demonstrates the ability of autonomous agents to perform simple tasks.
In order to solve complex task, the multi-agent framework has been widely explored by many researchers ~\cite{guo2024large}. 
CAMEL~\cite{li2023camel} and AutoGen~\cite{wu2023autogen} focuses on complex solutions through communication among agents.
ChatDev~\cite{qian2023communicative} and MetaGPT~\cite{hongmetagpt} split the process of program development into several stages that each engages an agent to facilitate a seamless workflow. The same strategy has been used in recommendation~\cite{wang2024multi, zhang2024generative}, debate~\cite{du2023improving,chan2023chateval}, question-answering ~\cite{sun2024harnessing} and fact-checking~\cite{kim2024can, liu2025bidev}. 
The multi-agent framework has also been applied to many social simulation works, where many role-played agents simulate the development of the society through the interaction and cooperation~\cite{park2023generative, zhang2023exploring, kaiya2023lyfe, liu2024skepticism, liu2024tiny}.
While the multi-agent framework on mobile scenarios is still under-explored. MobileAgent-v2~\cite{wang2024mobilev2} integrates planning, decision and reflection agents forming a pipeline equipped with memory unit to improve the performance of automated task execution, while it still struggle for cross-app instructions.

Most of the current multi agent frameworks use procedure-oriented agent splitting, while cross-app instructions are more suitable for object-oriented approach, thus we build an app-oriented multi-agent framework with self-evolution.

\section{MobileSteward}

\begin{figure*}[]
\setlength{\abovecaptionskip}{0.3cm}
\setlength{\belowcaptionskip}{-0.3cm}
\centering
\includegraphics[width=0.9\linewidth]{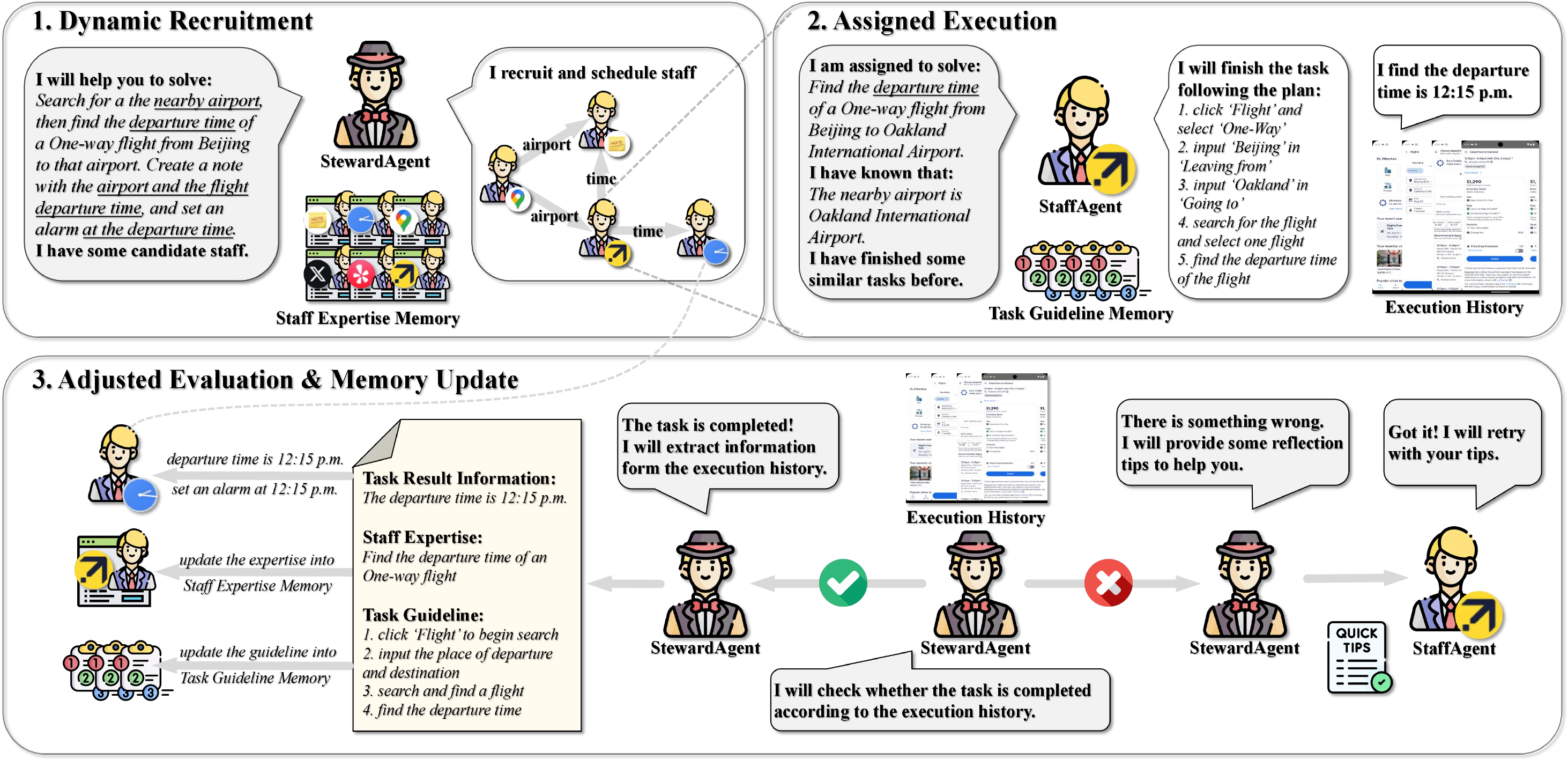}
\caption{MobileSteward consists of a centralized StewardAgent and several app-oriented StaffAgents. The framework integrates three modules: (1) \textit{Dynamic Recruitment}: StewardAgent splits the instruction into app-oriented tasks and generate StaffAgent scheduling graph.; (2) \textit{Assigned Execution}: StaffAgent automates the assigned task and returns the execution history; (3) \textit{Adjusted Evaluation}: StewardAgent provides reflection tips for wrong execution and summarizing successful executions to facilitate information transfer and adjust subsequent schedule.}
\label{fig:model}
\end{figure*}

\subsection{Task Formulation}

Mobile Task Automation is to automatically complete an instruction $I$ through an action sequence~\cite{wen2024autodroid}. At each step $i$, the model $\Phi$ decides the next action $a_{j,i}$ based on the current state information $S_{j,i}$ obtained from the mobile phone environment $E$. Thus, the instruction is automatically performed by an execution history $H$. 

While for complex cross-app instructions, which essentially require executing a sequence of tasks $T_j$ in the corresponding $App_j$, the execution history $H$ can be further detailed as follows:

\begin{align}
    I &= [T_1,\cdots,T_j,\cdots,T_m],\\
    H &= [H_1, \cdots, H_j, \cdots, H_m], \\
    H_j &= [a_{j,1}, \cdots, a_{j,i}, \cdots, a_{j,n}], \\
    a_{j,i} &= \Phi(S_i, T_j)
\end{align}

Therefore, to automatically execute complex cross-app instructions, it is essential to ensure: 
(1) \textit{Instructive task scheduling}, which involves decomposing instruction $I$ and scheduling the task $T_j$; 
(2) \textit{Accurate task execution}, which ensures the successful completion of task $T_j$ through precise action $a_{ji}$ within the $App_j$;
(3) \textit{Phased task adjustment}, which reflects on the error within the execution history $H_j$ and deliver key information between $H_j$.

\subsection{Framework Overview}

We propose \textbf{MobileSteward}, a self-evolving multi-agent framework, consists of a centralized StewardAgent and multiple app-oriented StaffAgents. As shown in Figure~\ref{fig:model}, we design three specialized modules within the MobileSteward: 
(1) \textbf{Dynamic Recruitment}: StewardAgent splits the instruction and schedules the corresponding StaffAgents. 
(2) \textbf{Assigned Execution}: StaffAgent executes the assigned task on the target app. 
(3) \textbf{Adjusted Evaluation}: StewardAgent evaluates the execution results, provides reflection, delivers information and adjusts the schedule. 
In order to improve the multiple agents within the framework, we equip the StewardAgent with a \textit{Staff Expertise Memory} for the cognition of StaffAgents' expertise, and equip the StaffAgents with a \textit{Task Guideline Memory} for task execution demonstrations. 
We provide a pseudo-code of MobileSteward in Algorithm~\ref{alg:algorithm} and we will detail the design of the multiple agents and the entire framework in subsequent sections.

\begin{algorithm}[t]
\caption{MobileSteward Working-Flow.}
\label{alg:algorithm}
\begin{flushleft}
\textbf{Input}: task instruction, $I$; mobile phone environment, $E$; max try times, $N_{try}$; max execution steps, $N_{step}$. \\
\textbf{Agents}: StewardAgent, $\Phi_{steward}$; StaffAgent, $\Phi_{staff}$. \\
\textbf{Memory}: staff expertise memory, $M_E$; task guideline memory, $M_G$. \\
\end{flushleft}
\begin{algorithmic}[1] 
\STATE \# \textbf{\textit{Dynamic Recruitment}}
\STATE $SG$ = Schedule($I,M_E;\Phi_{steward}$)
\FOR{$(T_j, \Phi_{staff_j})$ in Topological\_Sorting($SG$)}
\STATE try\_cnt = 0;\ $t_1$ = None
\WHILE{try\_cnt < $N_{try}$}
\STATE try\_cnt += 1
\STATE \# \textbf{\textit{Assigned Execution}}
\STATE step\_cnt = 0;\ $s_0$ = None;\ $H_j$ = []
\STATE $p_j$ = Plan($T_j,M_G;\Phi_{staff_j}$)
\WHILE{step\_cnt < $N_{step}$ \OR $a_{j,i}$ != FINISH}
\STATE step\_cnt += 1
\STATE $S_{j,i}$ = Get\_State($E$)
\STATE $a_{j,i}$ = Predict($T_j,S_{j,i},M_G,s_{j,i-1},p_j,R_j,t_j;\Phi_{staff_j}$)
\STATE $s_{j,i}$ = Summary($T_j,S_{j,i},a_{j,i};\Phi_{staff_j}$)
\STATE $E$ = Update\_State($E, a_{j,i}$)
\STATE $H_j$.append(($S_{j,i},a_{j,i},s_{j,i}$))
\ENDWHILE
\STATE \# \textbf{\textit{Adjusted Evaluation}}
\STATE $e_j$ = Evaluate($H_j,T_j;\Phi_{steward}$)
\IF{$e_j$ == ERROR}
\STATE $t_j$ = Reflect($H_j,T_j;\Phi_{steward}$)
\ELSE
\STATE $r_j,m_e,m_g$ = Extract($H_j,T_j;\Phi_{steward}$)
\FOR{($T_j,T_k$) in $SG$}
\STATE $T_k$ = Adjust($T_k,r_j;\Phi_{steward}$)
\STATE $R_k$.append($r_j$)
\ENDFOR
\STATE \# \textbf{\textit{Memory Update}}
\STATE $M_E$ = Update($M_E,m_e;\Phi_{steward}$)
\STATE $M_G$ = Update($M_G,(T_j,m_g);\Phi_{steward}$)
\ENDIF
\ENDWHILE
\ENDFOR
\end{algorithmic}
\end{algorithm}

\subsection{StewardAgent and StaffAgent}

In a large manor, there will be a steward to convey the host's orders, and several staff in charge of specific jobs. Following this pattern, we build StewardAgent and StaffAgent via role-playing.
We inject the definition of the role at the beginning of the prompt. We will describe their responsibilities next.

\vspace{0.3em}
\noindent \textbf{StewardAgent} $\Phi_{steward}$ is responsible for controlling the entire task execution process, including: (1) \textit{Schedule}: scheduling the StaffAgent and scheduling tasks; (2) \textit{Evaluate}: evaluating the StaffAgent's task execution; (3) \textit{Reflect}: providing reflection and suggestions on wrong execution; (4) \textit{Extract}: extracting results and experience from successful execution; (5) \textit{Adjust}: delivering the key information and adjusting the subsequent schedule. (6) \textit{Update}: updating the memory for improvement.

\noindent \textbf{StaffAgent} $\Phi_{staff}$ is responsible for operating a specific app. In order to distinguish between different StaffAgents, we emphasize its proficiency in that app and add a app description in the role-playing prompt. Moreover, we equip StaffAgents with an app-specific Task Guideline Memory. The core task of StaffAgent is to execute the tasks in the app, including:
(1) \textit{Plan}: planning with the successful task execution; (2) \textit{Predict}: predicting the next action; 
(3) \textit{Summary}: summarizing the previous action. 
We adopt the AppAgent~\cite{yang2023appagent} to build StaffAgent, simplifying the action space as shown in the Table~\ref{tab:action_space}. We extract the interactive widgets and textual content from the XML, perform a hierarchical simplification. Then we mark these elements on the screenshot and feed the XML information aligned with the screenshot into the StaffAgent to predict the action. 

\begin{table}[]
\setlength{\abovecaptionskip}{0.3cm}
\caption{Action space used in StaffAgent}
\label{tab:action_space}
\begin{tabular}{l|l|l}
\hline \hline
\textbf{Type} & \textbf{Param}     & \textbf{Description}                      \\ \hline 
click         & element id         & click on the element with id     \\
input         & text content       & input the text content  \\
swipe         & up/down/right/left & swipe to some direction        \\
back          &  -                  & return to the previous page               \\
FINISH        &   -                 & finish the task                     \\ \hline \hline
\end{tabular}
\end{table}

\subsection{Dynamic Recruitment}

Complex cross-app instructions require scheduling of multiple StaffAgents to operate the apps. While, the scheduling of StaffAgents is dynamically aligned with the user instructions. Moreover, cross-app instructions often contain complex task association and information transfer between apps. To address these issues, we design Dynamic Recruitment to instruct StewardAgent to generate a scheduling graph for StaffAgents with the guidance of information flow. We also equip the StewardAgent with a Staff Expertise Memory that records the app description and expertise list.

On receiving the instruction $I$, StewardAgent decomposes the instruction into sub-tasks on the specific apps based on Staff Expertise Memory $M_E$, and then analyzes the information flow between these tasks, and outputs these contents as \textit{thought}. Subsequently, based on the previous thought, StewardAgent recruits the StaffAgents corresponding to these apps and constructs the scheduling graph $SG$ among them, which is exported as \textit{plan}. The process can be described as:

\begin{equation}
    SG = \text{Schedule}(I, M_E;\Phi_{steward})
\end{equation}

\subsection{Assigned Execution}

Due to the significant variance in functionality, content, and layout between apps, we dedicate each StaffAgent to operate a specific app, and equip each StaffAgent with an app-specific Task Guideline Memory to support its task execution. The scheduling graph generated in the Dynamic Recruitment phase is a DAG, so we use topological sorting on the scheduling graph to assign tasks to the corresponding StaffAgent $\Phi_{staff_j}$ for execution. 
  
StaffAgent first extracts the successful execution demonstrations related to the assigned task $T_j$ from memory $M_G$ to make a task plan $p_j$. At each execution step $i$, StaffAgent obtains the current state information $S_{j,i}$ from the mobile phone environment $E$, which contains the screenshots and XML layout file. Combining the received result information $R_j$ and reflection tips $t_j$ obtained from the preceding execution, guided by the task plan $p_j$, StaffAgent will decide an action $a_{j,i}$ to advance the assigned task $T_j$ on the current state $S_{j,i}$. After each execution, StaffAgent will generate the current action summarization $s_{j,i}$, which will conclude the previous action sequence, the current execution result and the functional description of the related GUI element. 
After completing the task or reaching the maximum number of steps, the StaffAgent packages the execution history $H_j$. We can formulate the process as follows:
\vspace{-5pt}
\begin{align}
    p_j &= \text{Plan}(T_j, M_G;\Phi_{staff_j}) \\
    a_{j,i} &= \text{Predict}(T_j, S_{j,i}, M_G, s_{j,i-1}, p_j, R_j, t_j;\Phi_{staff_j}) \\
    s_{j,i} &= \text{Summary}(T_j, S_{j,i}, a_{j,i};\Phi_{staff_j}) \\
    H_j &= [(S_{j,1}, a_{j,1}, s_{j,1}),\cdots,(S_{j,n}, a_{j,n}, s_{j,n})]
\end{align}
\vspace{-15pt}
\subsection{Adjusted Evaluation}

In order to better control the execution of tasks and the advancement of scheduling, we design Adjusted Evaluation which utilizes the StewardAgent to evaluate the execution process of the StaffAgents, providing reflection tips on error execution or summarizing the correct execution to deliver key information and adjust succeeding schedule according to the scheduling graph.

StewardAgent will generate the evaluation $e_j$ on the simplified execution history $H_j$ of assigned task $T_j$ from StaffAgent $\Phi_{staff_j}$. 
If there exist errors, StewardAgent will give reflection tips $t_j$ that will contribute to the task automation. 
If the task is completed, StewardAgent will extract the task result information $r_j$, the staff expertise $m_e$ and the task guideline $m_g$ from the execution process. Subsequently, the task result information will be delivered according to the scheduling graph $SG$, and used to adjust the succeeding task schedules. The Adjusted Evaluation will be formulated as:
\vspace{-5pt}
\begin{align}
    e_j &= \text{Evaluate}(H_j, T_j; \Phi_{steward}) \\
    t_j &= \text{Reflect}(H_j, T_j; \Phi_{steward}),if\ e_j == \text{ERROR} \\
    r_j, m_e, m_g &= \text{Extract}(H_j, T_j; \Phi_{steward}),if\ e_j == \text{SUCCESS} \\
    T_k &= \text{Adjust}(T_k, r_j; \Phi_{steward}), if\ (T_j, T_k) \in SG \\
    R_k & .\text{append}(r_j), if\ (T_j, T_k) \in SG
\end{align}
\vspace{-15pt}
\subsection{Memory-Based Self-Evolution}

As in reality, with the accumulation of experience, the steward will become more aware of the staff's suitable job and staff will become more proficient at their job. Therefore, we propose Memory-Based Self-Evolution for continuous improvement of MobileSteward. Specifically, we equip the StewardAgent with a Staff Expertise Memory records a description of the app as well as a list of the expertise of the StaffAgent. These will be used for task decomposition and scheduling during the Dynamic Recruitment. Meanwhile, we equip the StaffAgent with Task Guideline Memory, which records successful task steps. These demonstrations will be used as references for planning and predicting in the Assigned Execution.

MobileSteward's self-evolution is achieved by constantly updating both memories. After the StaffAgent has successfully completed the assigned task, StewardAgent will extract the staff expertise $m_e$ and task guideline $m_g$, and then use them to update the Staff Expertise Memory $M_E$ and Task Guideline Memory $M_G$ respectively. When updating Staff Expertise Memory, StewardAgent needs to determine whether the newly extracted $m_e$ needs to be updated into the memory. When updating the Task Guideline Memory, the StewardAgent updates the ($T_j$,$m_g$) pairs into the Memory. The process of memory update can be described as follows:
\vspace{-5pt}
\begin{align}
    M_E &= \text{Update}(M_E, m_e; \Phi_{steward}) \\
    M_G &= \text{Update}(M_G, (T_j, m_g); \Phi_{steward})
\end{align}
\vspace{-15pt}

\begin{table*}[]
\setlength{\abovecaptionskip}{0.3cm}
\renewcommand\arraystretch{1.4}
\caption{The overall performance of baselines and MobileSteward on both CAPBench and SAPBench. We list the base model used in all of the methods for a clearer comparison and implement MobileSteward with both GPT-4v and GPT-4o.}
\begin{tabular}{l|l|ccc|ccc}
\hline \hline
\multirow{2}{*}{\textbf{Method}} & \multicolumn{1}{l|}{\multirow{2}{*}{\textbf{Model}}} & \multicolumn{3}{c|}{\textbf{CAPBench}}        & \multicolumn{3}{c}{\textbf{SAPBench}}      \\ \cline{3-8} 
                                 & \multicolumn{1}{l|}{}                                    & Success Rate & Task Rate & App Rate & Success Rate & Step Rate & App Rate \\ \hline
AutoDroid~\cite{wen2024autodroid}                         & GPT-4                                                   & 0.00          & 0.10         & 0.31           & 0.28          & 0.35      & 0.84           \\
AppAgent~\cite{yang2023appagent}                         & GPT-4v                                                   & 0.00          & 0.11         & 0.31           & 0.50          & 0.62      & 0.88           \\
MobileAgent~\cite{wang2024mobile}                   & GPT-4v                                                   & 0.01          & 0.11         & 0.33           & 0.38          & 0.57      & 0.80           \\
MobileAgent-v2~\cite{wang2024mobilev2}                   & GPT-4o                                                   & 0.21          & 0.37         & 0.67           & 0.64          & 0.74      & 0.98           \\ \hline
MobileSteward                    & GPT-4v                                                   & 0.55          & 0.76         & 0.99           & 0.76          & 0.81      & 0.98           \\
MobileSteward                    & GPT-4o                                                   & 0.59          & 0.79         & 1.00           & 0.78          & 0.87      & 1.00           \\ \hline \hline
\end{tabular}
\label{tab:experiment}
\end{table*}

\section{Experiments}

\subsection{Benchmarks}

The evaluation of mobile phone agents is more convincing in a real-world interactive environment, which is more complex and dynamic~\cite{deng2024mobile}. Therefore, we built a simulation environment with Pixel 8 Pro in Android Studio and used ADB to complete the interaction with the simulator. We conduct evaluation on two benchmarks.

\begin{figure}[H]
	\centering
	\begin{minipage}{0.44\linewidth}
		\centering
		\includegraphics[width=\linewidth]{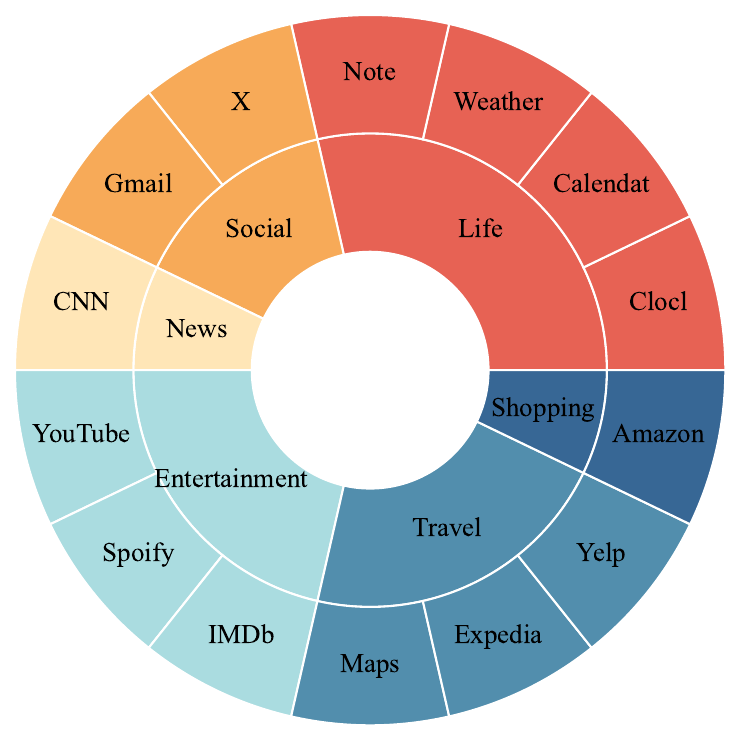}
		\caption{App categories}
		\label{fig:app_categories}
	\end{minipage}
	\begin{minipage}{0.54\linewidth}
		\centering
		\includegraphics[width=\linewidth]{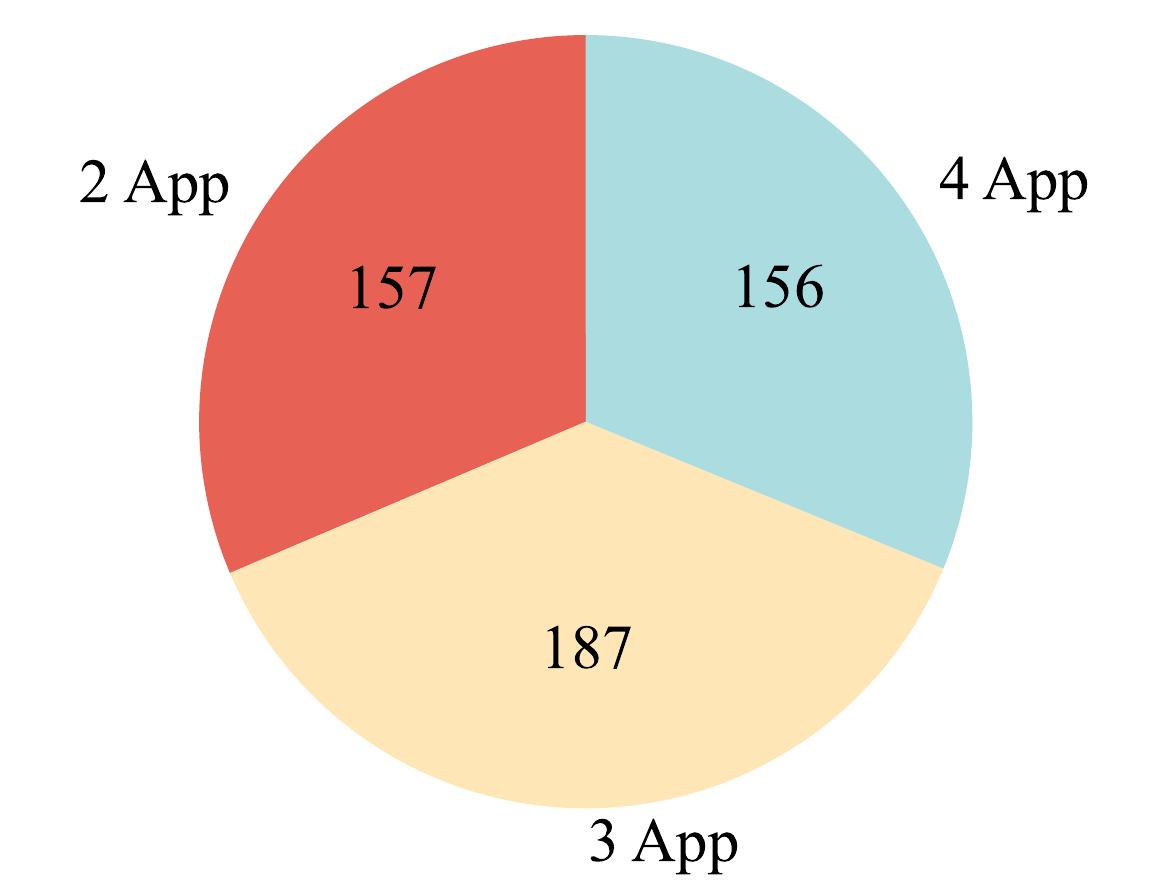}
		\caption{Task Statistics}
		\label{fig:task_number}
	\end{minipage}
\end{figure}

\vspace{0.5em}
\noindent \textbf{Cross-APP Benchmark}:
We construct \textbf{CAPBench}, which is specialized for complex cross-app instructions. CAPBench takes into full consideration about the task association and information transfer that exists among apps in real-world scenarios. We select a total of 14 apps in 6 categories, including life, social, news, entertainment, shopping and traveling, and ensure that there exist reasonable task associations among these apps. To generate the cross-app instruction data, we manually annotated each app with a function description as well as a list of common task templates. When constructing the data, we randomly select 2-4 apps from the candidate apps, and then prompt GPT-4 with the app information to analyze the existence of reasonable task associations and information transfer among the apps, and then select the corresponding task templates to be instantiated to assemble the cross-app instructions. We then ask the annotators to perform a human evaluation, eliminating invalid instructions. In total, we constructed 500 cross-app instructions, we present a statistics for app categories in Figure~\ref{fig:app_categories} and the number of tasks by complexity in Figure~\ref{fig:task_number}.

\vspace{0.5em}
\noindent \textbf{Single-APP Benchmark}:
For more comprehensive evaluation of our framework, we collect \textbf{SAPBench} from the previous works~\cite{yang2023appagent,wang2024mobile} with a total of 50 instructions.

\subsection{Experimental Setup}

\subsubsection{Baselines}

To verify the effectiveness of our proposed framework, we compare MobileSteward with both single agent and multi agent baselines.

\noindent (1) \textbf{AppAgent}~\cite{yang2023appagent} uses XML file to extract interactive elements and generates element-level function documents by self-exploration and demo-watching, which will be used to assist action decision. We add a \textit{home()} action to complete the cross-app instructions.

\noindent (2) \textbf{MobileAgent}~\cite{wang2024mobile} introduces a visual perception module to localize the natural language described actions on the screen.

\noindent (3) \textbf{MobileAgent-v2}~\cite{wang2024mobilev2} proposes a process-oriented multi-agent framework that integrates planning, decision and reflections agents to form a working-flow, which is equipped with a memory unit to store and retrieve key information during execution. 

\subsubsection{Evaluation Metrics}

We design a multi-granularity evaluation metric, including success rate and app rate. For cross-app instructions, we use a app-level task rate, while for single-app instructions, we use a more fine-grained step rate.

\noindent $\bullet$ \textit{Success Rate}: To evaluate whether the instruction is completed.

\noindent $\bullet$ \textit{App Rate}: To evaluate the percentage of the overlap between the apps covered in the execution and the labeled apps.

\noindent $\bullet$ \textit{Task Rate}: To evaluate the app-level completion of instruction, which is the ratio of the apps completing the task to the total apps.

\noindent $\bullet$ \textit{Step Rate}: To evaluate the step-level action accuracy, which is the ratio of the correct steps to the total steps during execution.

\subsubsection{Implementation Details}

In our proposed MobileSteward, we use the same base model to build StewardAgent and StaffAgents, that we used the \textit{gpt-4-vision-preview} version of GPT-4v, and the \textit{gpt-4o (2024-05-13)} version of GPT-4o, and we set temperature to 0.
For Dynamic Recruitment and Adjusted Evaluation, we prompt with a 2-shot in-context learning. For Assigned Execution, we use zero-shot prompting. We set the max try $N_{try}$ to 3 and max step $N_{step}$ to 20 for MobileSteward, and set max step $N_{step}$ to 80 for all of the baselines. 
StaffAgent uses BM25~\cite{robertson2009probabilistic} to retrieve top-3 most relevant tasks from Task Guideline Memory for reference.
For Self-Evolution, we sample 50 instructions from each complexity split of CAPBench for test and use the remaining instructions for prior self-evolution, we also update the memories during the test. 
We build up a mobile phone environment with the Pixel 8 Pro in Android Studio. We use the API level of 34 and the Target of Android 14 (Google Play)  \footnote{Code will be available at: \url{https://github.com/XiaoMi/MobileSteward}}.

\subsection{Overall Performance}

We compared MobileSteward with all baselines on two benchmarks and the experimental results are shown in the Table~\ref{tab:experiment}. The experimental results demonstrate the following conclusions: 

\vspace{0.1cm}

\noindent \textbf{MobileSteward can effectively solve both single and cross app instructions.}
The experimental results for Success Rate indicate that baseline methods perform terribly on CAPBench. Both AppAgent and MobileAgent struggle to complete cross-app instructions, with their App Rate being 60\% lower than that of MobileSteward. This suggests they have difficulty in selecting the appropriate apps to accomplish the instruction. The Task Rate reflects that the long execution sequence leads to error propagation and information loss. Thus the best performance of MobileSteward demonstrate that Staff Expertise Memory contributes to assigning task to appropriate app-oriented StaffAgent in Dynamic Recruitment. Furthermore, Adjusted Evaluation can effectively alleviate error propagation and information loss between StaffAgents.
Although MobileSteward is designed to address complex cross-app instructions, experimental results show that it is equally effective to solve simple single-app instructions. The baseline methods is much lower on App Rate, which demonstrates that our self-evolution of Staff Expertise Memory can effectively improve the schedule of StaffAgent.

\vspace{0.1cm}

\noindent \textbf{App-oriented multi-agent framework is more effective.} 
From the experimental results, MobileAgent-v2 and MobileSteward perform better on both cross app and single app instructions, indicating that the multi-agent methods is more effective compared to the single-agent methods. 
Compared to procedure-oriented MobileAgent-v2, MobileSteward is 38\% and 14\% higher on CAPBench and SAPBench respectively. The great diversity between apps result in a 30\% lower Task Rate of MobileAgent-v2 because it is difficult to use one agent to handle all of apps. 
While, our proposed MobileSteward is an app-oriented multi-agent framework, and the experimental results demonstrate that Dynamic Recruitment and Assigned Execution can schedule more appropriate app-oriented StaffAgents to execute the assigned task effectively. Meanwhile, unlike MobileAgent-v2, which is a static framework, our proposed Memory-based Self-evolution mechanism can dynamically improve the StaffAgent's expertise in the specific app and the StewardAgent's schedule of the tasks.
Therefore, our app-oriented multi-agent framework is more effective to solve the cross-app instructions.

\begin{table}[]
\setlength{\abovecaptionskip}{0.3cm}
\caption{Ablation Study for Self-Evolution, Assigned Execution and Adjusted Evaluation.}
\renewcommand\arraystretch{1.3}
\setlength{\tabcolsep}{3pt}
\begin{tabular}{l|ccc}
\hline \hline
\textbf{Method}         & \textbf{Success Rate} & \textbf{Task Rate} & \textbf{App Rate} \\ \hline
MobileSteward           & 0.55          & 0.81         & 0.99           \\
w/o Self-Evolution      & 0.50          & 0.77         & 0.94           \\
w/o Assigned Execution  & 0.44          & 0.71         & 0.98           \\
w/o Adjusted Evaluation & 0.36          & 0.68         & 0.94           \\ \hline \hline
\end{tabular}
\label{tab:ablation}
\end{table}

\subsection{Ablation Study}

We design ablation experiment to fully explore the effectiveness of our proposed modules, the experimental results are shown in Table~\ref{tab:ablation}.
We can find that: (1) When Self-Evolution is disabled, we find that both Task Rate and App Rate decreased. which demonstrate that in Self-Evolution, StaffAgent improves task execution through Task Guideline Memory, and StewardAgent improves task schedule between apps through Staff Expertise Memory.
(2) When Assigned Execution is removed, the decrease of Success Rate and Task Rate is more obvious, which indicates that using app-oriented StaffAgent can effectively improve the execution of tasks because they have more guideline information of the specific app.
(3) The absence of Adjusted Evaluation causes the decrease in App Rate, which illustrates its impact on Self-Evolution. The decreases in Success Rate and Task Rate indicate that Adjusted Evaluation is effective in evaluating the execution process and can provide reflective suggestions to correct incorrect execution.

\subsection{Further Analysis}

\subsubsection{Analysis on MobileAgentBench}


We evaluate our proposed MobileSteward on MobileAgentBench~\cite{wang2024mobileagentbench}, which consists of 100 tasks across 10 simple system apps. To assess performance, we employ five metrics: 
(1) SR: Success Rate;
(2) SE: Step-wise Efficiency;
(3) IOT: Input-Output Tokens;
(4) FN: False Negative;
(5) FP: False Positive.
The experimental results, as shown in Table~\ref{tab:mobileagentbench}, demonstrate that MobileSteward achieves the highest SR while maintaining competitive SE, indicating its efficiency in completing simple tasks. The slightly higher FN can be attributed to the smaller number of failed task samples; however, this value normalizes to 0.16 when adjusted for the total number of task samples. Additionally, the lower FP suggests that MobileSteward makes more accurate “FINISH” decisions upon reaching the final state, further showcasing its robustness.

\begin{table}[]
\setlength{\abovecaptionskip}{0.3cm}
\renewcommand\arraystretch{1.3}
\setlength{\tabcolsep}{5pt}
\caption{Experimental results on MobileAgentBench and the results with $^\dag$ are reported by~\cite{wang2024mobileagentbench}.}
\begin{tabular}{l|ccccc}
\hline \hline
              & \textbf{SR} & \textbf{SE} & \textbf{IOT} & \textbf{FN} & \textbf{FP} \\ \hline
AndroidArena$^\dag$~\cite{xing2024understanding}  & 0.22        & 1.13        & 780.47          & 0.09        & 0.33        \\
AutoDroid$^\dag$~\cite{wen2024autodroid}     & 0.27        & 3.10        & 963.48          & 0.93        & 0.01        \\
CogAgent$^\dag$~\cite{hong2024cogagent}      & 0.08        & 2.42        & 579.84          & 1.0         & 0.04        \\
AppAgent$^\dag$~\cite{yang2023appagent}      & 0.40        & 1.29        & 1505.09         & 0.17        & 0.40        \\
MobileAgent$^\dag$~\cite{wang2024mobile}   & 0.26        & 1.13        & 1236.88         & 0.19        & 0.31        \\ \hline
MobileSteward & 0.58        & 1.24        & 2249.09         & 0.40        & 0.11        \\ \hline \hline
\end{tabular}
\label{tab:mobileagentbench}
\end{table}

\subsubsection{Analysis of Complexity on CAPBench}

\begin{figure}[]
\setlength{\abovecaptionskip}{0.3cm}
\setlength{\belowcaptionskip}{-0.4cm}
\centering
\includegraphics[width=0.75\linewidth]{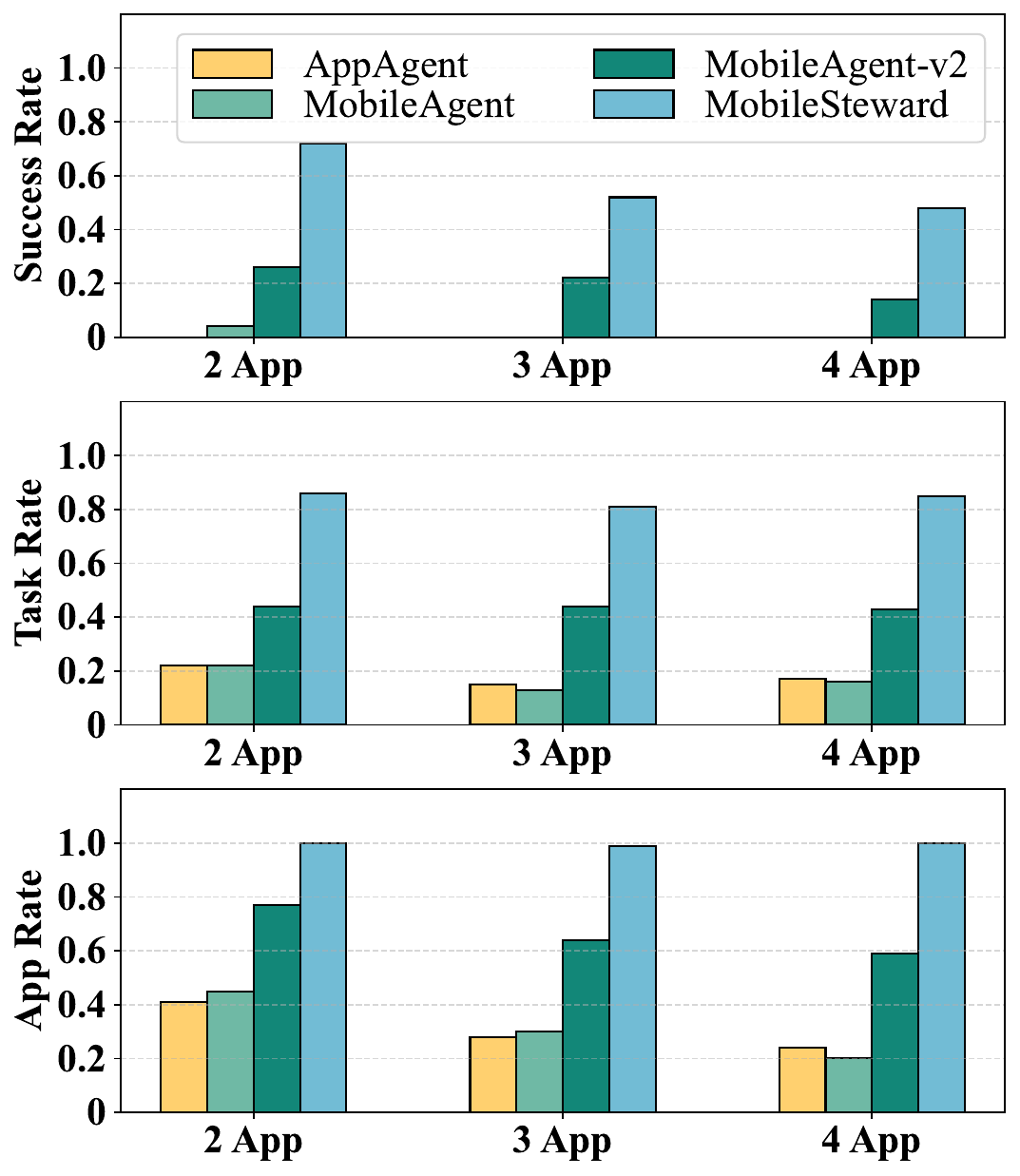}
\caption{Complexity analysis on CAPBench.}
\label{fig:complexity}
\end{figure}

We conducted a more detailed in-depth analysis of CAPBench using the number of apps involved in the task as a measure of complexity. The analysis results are shown in Figure~\ref{fig:complexity} . As the task complexity increases, the Success Rate of all methods decreases dramatically. MobileAgent directly fails to complete the task with 4 apps. And MobileAgent-v2 also decreased by 46\% from 2app to 4app. In comparison, our MobileSteward only decreased by 33\% as the task complexity increased. And compared to MobileAgent-v2, we improve 1.77 times on 2app and 2.43 times on 4app, which all proves that our MobileSteward performs better on more complex cross-app instruction. Meanwhile, as for the App Rate, MobileSteward remains stable as the complexity rises, while all of the baselines decrease to different degrees, indicating that MobileSteward is able to accomplish the scheduling between tasks more efficiently.

\begin{table}[]
\setlength{\abovecaptionskip}{0.3cm}
\setlength{\belowdisplayskip}{1pt}
\renewcommand\arraystretch{1.3}
\setlength{\tabcolsep}{4pt}
\caption{Analysis of Dynamic Recruitment. Bold numbers mean the best performance using a closed-source base model, and underlined numbers indicate the best performance using an open-source base model.}
\begin{tabular}{l|l|c|ccc}
\hline \hline
\textbf{Method}                         & \textbf{Model} & \textbf{\#Param} & \textbf{2App} & \textbf{3App} & \textbf{4App} \\ \hline
MobileAgent-v2                 & GPT-4o         & -       & 0.26 & 0.22 & 0.14                     \\ \hline
\multirow{5}{*}{MobileSteward} & Qwen-Vl        & 9.6B    & 0.28 & 0.12 & 0.06                     \\
                               & GLM-4V         & 14B     & 0.36 & 0.22 & 0.16                     \\
                               & InternVL       & 26B     & 0.40 & 0.20 & 0.20                     \\
                               & InternVL2      & 40B     & \underline{0.50} & \underline{0.30} & \underline{0.22}                     \\
                               & GPT-4o         & -       & \textbf{0.72} & \textbf{0.58} & \textbf{0.48}                     \\ \hline \hline
\end{tabular}
\label{tab:dynamic_recruitment}
\end{table}

\subsubsection{Analysis of Dynamic Recruitment}

In order to validate the effectiveness of our Dynamic Recruitment in task scheduling, we use different base models that have difference numbers of parameters and capabilities to accomplish the Dynamic Recruitment. The comparison results are shown in Table~\ref{tab:dynamic_recruitment}. Compared to the naive text plan used in MobileAgent-v2, our proposed Dynamic Recruitment is more effective, that we can outperform MobileAgent-v2 equipped with GPT-4o using only a 14B GLM-4V. Because we use information flow to guide the generation of the scheduling graph between StaffAgents, which can explicitly establish the association between tasks, including the scheduling order and information transfer between them. This contains more information than a naive text plan, and therefore gives clearer guidance during subsequent execution and ensures the efficient transfer of information.

\subsubsection{Analysis of Self-Evolution}

\begin{figure}[]
\setlength{\abovecaptionskip}{0.3cm}
\setlength{\belowcaptionskip}{-0.4cm}
  \centering
  \includegraphics[width=0.75\linewidth]{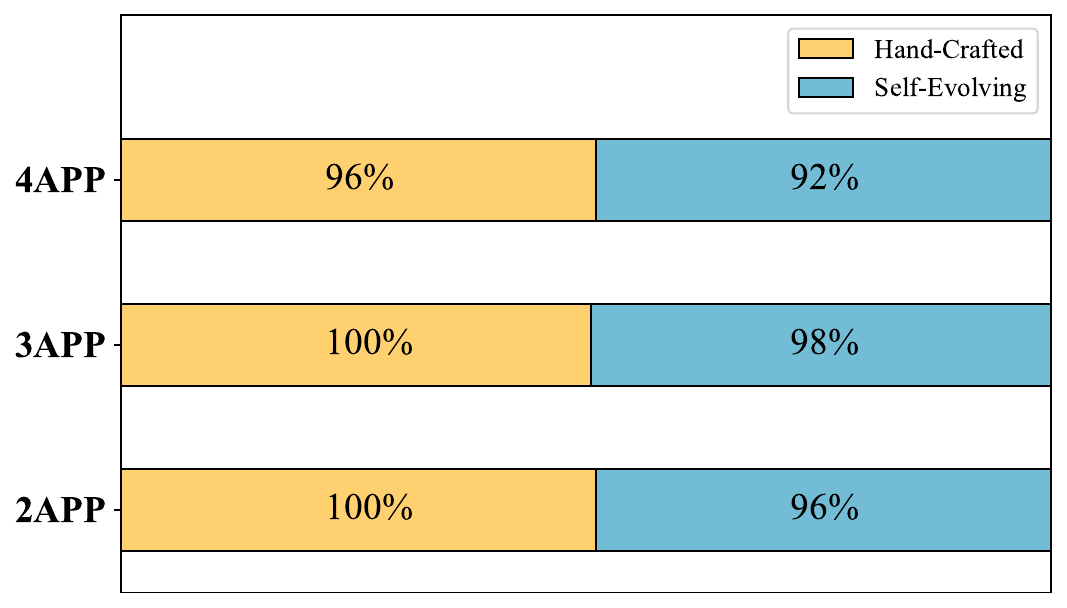}
  \caption{Comparison of hand-crafted staff expertise and self-evolving staff expertise.}
  \label{fig:self_evolution}
  \vspace{-2pt}
\end{figure}

In order to validate the effectiveness of our proposed self-evolution, we have designed experiments to compare the accuracy of task scheduling using hand-crafted and self-evolving staff expertise. As shown in Figure~\ref{fig:self_evolution}, the self-evolving staff expertise can achieve comparable results with hand-crafted, which validates the effectiveness of our proposed self-evolution that it can summarize the staff expertise from the successful execution and assist in the task scheduling.

\begin{figure*}[]
\setlength{\abovecaptionskip}{0.3cm}
\setlength{\belowcaptionskip}{-0.3cm}
  \centering
  \includegraphics[width=0.85\linewidth]{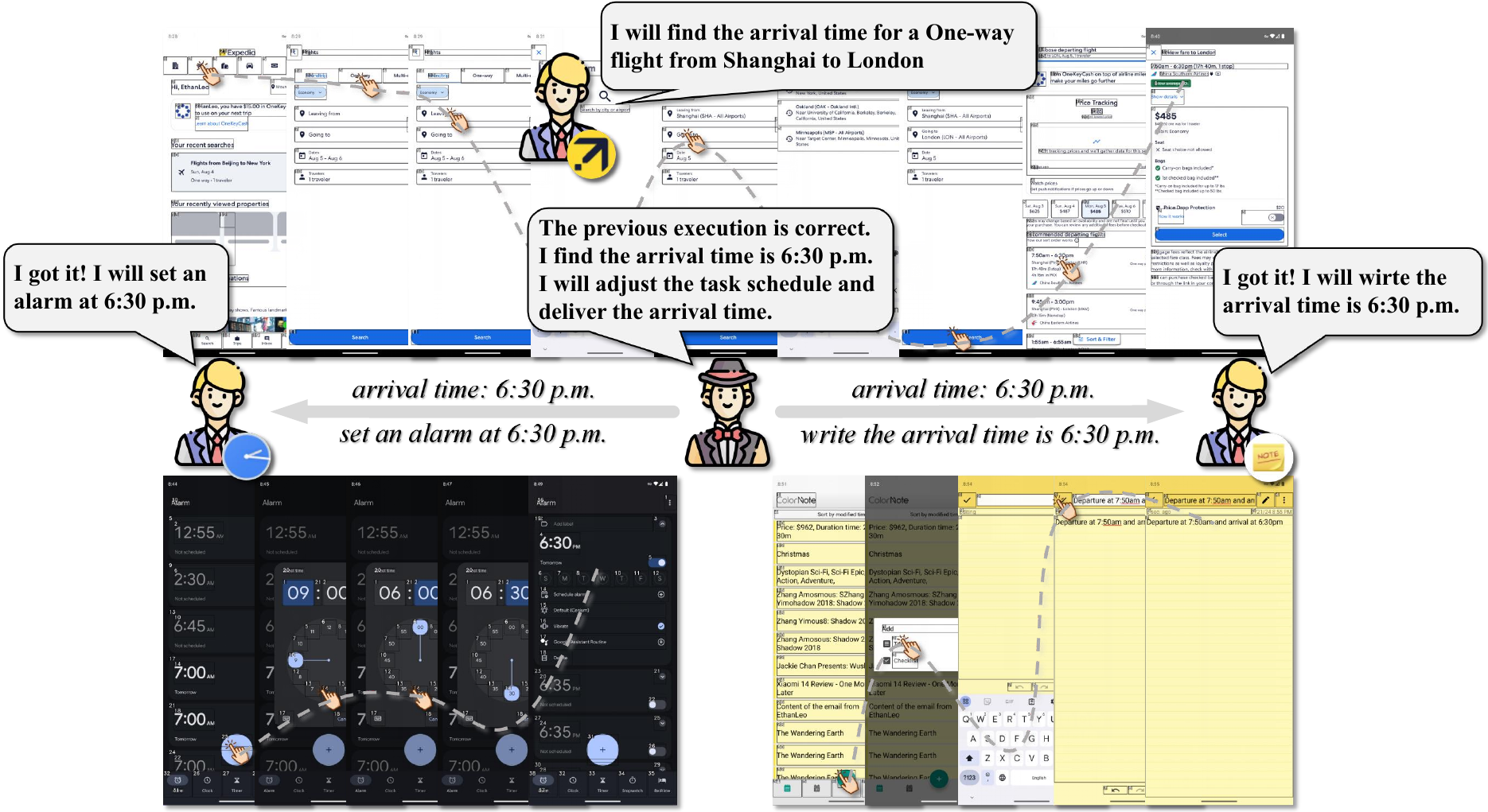}
  \caption{Case study for the task: \textit{Search for a One-way flight(From Shanghai to London), set an alarm for the arrival time, and create a note with the flight information}. We illustrate the process of Assigned Execution and Adjusted Evaluation.}
  \label{fig:case}
\end{figure*}

\subsubsection{Analysis of Efficiency}

We compare our proposed MobileSteward with the strong baseline MobileAgent-v2 to evaluate efficiency. To ensure a comprehensive assessment, we utilize two metrics:
(1) Actions per Task (A/P): The number of actions required to complete a task.
(2) Tokens per Action (T/A): The number of tokens consumed for each action decision.
As shown in Table~\ref{tab:efficiency}, while MobileAgent-v2 demonstrates comparable efficiency to MobileSteward on simple tasks (SAPBench), it requires more actions to complete complex tasks (CAPBench). Moreover, MobileSteward consumes fewer tokens per action, demonstrating its ability to achieve superior results with lower computational cost and higher overall efficiency.

\subsubsection{Analysis of Online User Experiments}

We evaluate MobileSteward on 50 tasks provided by 10 online users, with the experimental results presented in Table~\ref{tab:online}. MobileSteward demonstrates comparable performance in Success Rate and Task Rate across both online and offline environments. The observed decline in App Rate can be attributed to the ambiguity in user instructions, which occasionally prevents the system from identifying the exact target application.

\begin{table}[]
\setlength{\abovecaptionskip}{0.2cm}
\renewcommand\arraystretch{1.3}
\caption{Analysis of Efficiency. A/T represents Actions per Task; T/A represents Tokens per Action.}
\begin{tabular}{l|cc|c}
\hline \hline
               & \multicolumn{2}{c|}{\textbf{A/T}} & \multicolumn{1}{c}{\textbf{T/A}} \\
               & \textbf{CAPBench}        & \textbf{SAPBench}        & \textbf{Total AVG}                        \\ \hline
MobileAgent-v2 & 39.60           & 11.52           & 3194.23                          \\
MobileSteward  & 22.06           & 9.13            & 2687.37                          \\ \hline \hline
\end{tabular}
\label{tab:efficiency}
\vspace{-0.3cm}
\end{table}

\subsection{Case Study}

We illustrate the execution of an instruction with a complexity level of 3App in Figure~\ref{fig:case}, including the Assigned Execution of StaffAgent and Adjusted Evaluation of StewardAgent. After StaffAgent specializing in Expedia finds the arrival time of a flight from Shanghai to London, StewardAgent evaluates that the execution has successfully completed the task and extracts the task result information: \textit{the arrival time is 6:30 p.m.}. Then StewardAgent delivers the task result information to the StaffAgent specialized in Clock and Note based on the scheduling graph generated in Dynamic Recruitment and adjusts the task assigned to them with the task result information.

\begin{table}[]
\setlength{\abovecaptionskip}{0.2cm}
\renewcommand\arraystretch{1.3}
\caption{Analysis of online user experiments.}
\begin{tabular}{l|ccc}
\hline \hline
        & \textbf{Success Rate} & \textbf{Task Rate} & \textbf{App Rate} \\ \hline
$\text{MobileSteward}_{\text{off}}$ & 0.59                  & 0.79              & 1.00               \\
$\text{MobileSteward}_{\text{on}}$  & 0.54                  & 0.76              & 0.87               \\ \hline \hline
\end{tabular}
\label{tab:online}
\vspace{-0.3cm}
\end{table}

\section{Conclusion}

We integrate multiple app-oriented StaffAgents coordinated by a centralized StewardAgent to constitute a self-evolving multi-agent framework named \textbf{MobileSteward}. For better executing cross-app instructions, we design three specific modules: \textit{Dynamic Recruitment} generates a scheduling graph to explicitly associate tasks among apps; \textit{Assigned Execution} assigns the task to an app-oriented StaffAgent to prevent the interference of diversity between apps; \textit{Adjusted Evaluation} conducts evaluation to alleviates error propagation or information loss during multi-step execution. We optimize MobileSteward using a \textit{Memory-base Self-evolution} mechanism that can learn from the successful execution. In order to evaluate our MobileSteward, we construct the first English Cross-APP Benchmark(CAPBench) in the real-world environment. The experimental results demonstrate that our MobileSteward achieve the best performance compared to both single-agent and multi-agent baselines on solving cross-app instructions.

\begin{acks}
This work was supported by the National Natural Science Foundation of China (NSFC Grant No. 62122089),
Beijing Outstanding Young Scientist Program NO. BJJWZYJH012019100020098, and Intelligent Social Governance Platform, Major Innovation \& Planning Interdisciplinary Platform for the “Double-First Class” Initiative, Renmin University of China, and the Research Fund of Xiaomi.
\end{acks}

\bibliographystyle{ACM-Reference-Format}
\bibliography{MobileSteward}

\appendix

\end{document}